\documentstyle[psfig]{mn}

\input{psfig.sty}

\newif\ifAMStwofonts

\title[Spectral and temporal variability of Seyfert 1s]
        {Thunderclouds and accretion discs: a model for the spectral and temporal variability of Seyfert 1 galaxies
}
\author[A. Merloni \& A. C. Fabian]
        { A. Merloni and A. C. Fabian
\\Institute of Astronomy, Madingley Road, Cambridge, CB3 0HA
}

\date{}

\begin{document}

\maketitle

\label{firstpage}

\begin{abstract}
X-ray observations of Seyfert 1 galaxies offer the unique possibility of observing spectral variability
on timescales comparable to the dynamical time of the inner accretion flow.
They typically show highly variable lightcurves, 
on a wide range of timescales, with Power Density Spectra characterized by `red noise' 
and a break at low frequencies. On the other hand, time resolved spectral analysis 
have established that spectral variability on the shortest timescales 
is important in all these sources, with the spectra getting 
softer at high fluxes (in the 2-10 keV band, typically), while the reflection component and 
the iron line often exhibit a complex behaviour. 
Here we present a model that is able to explain a number of the above mentioned properties in terms of 
magnetic flares shining above a standard accretion disc and producing the X-ray spectrum via
inverse Compton scattering of soft photons 
(both intrinsic and reprocessed thermal emission from the accretion disc 
and locally produced synchrotron radiation).
We show that the fundamental heating event, likely caused by magnetic reconnection,
must be compact, with typical size comparable to the accretion
disc thickness and must be triggered at a height at least an order of magnitude larger than
its size. 
The fundamental property of our `thundercloud' model is that 
the spatial and temporal distribution of flares are not random: the heating of the corona
 proceed in correlated trains of events in an avalanche fashion.
The  amplitude of the avalanches obeys a power-law 
distribution and determines the size of the active regions where the spectrum is produced. Due to the feedback effect of the X-ray radiation reprocessed 
in the disc, larger active regions produce softer spectra. With our model 
we simulate X-ray lightcurves that reproduce the main observational 
properties of the Power Density Spectra and of the 
X-ray continuum short-term variability of Seyfert 1 galaxies. 
By comparing them with observations of MGC--6-30-15, we are able
to infer that the accretion disc corona in this source must have a large optical depth ($\tau_{\rm T}\ga 1.5$) and
small average covering fraction.  
\end{abstract}


\section{Introduction}

X-ray observations of accreting black holes, both in Seyfert 1 galaxies and in
galactic sources (GBHC) typically show highly variable
lightcurves, on a wide range of timescales. Time resolved spectral analysis have established that
spectral variability on the shortest timescales is important in all these sources. For example,
the analysis of average peak aligned shots of the GBHC Cygnus X-1
(Li, Feng \& Chen 1999) has shown that, during the {\it hard} state
and the transitions between the two states of the source,
the spectrum evolves from hard to soft to hard again.
Such kind of analysis, though, is much better suited for Active Galactic Nuclei (AGN).
As a matter of fact, the dynamical timescale of a Keplerian accretion flow around a black hole,
$t_{\rm dyn} \simeq 9\times 10^3 \varpi^{3/2} (M_{\rm BH}/10^7 M_{\odot})$ s (where we
have introduced the dimensionless radius $\varpi=R/R_{\rm S}=c^2 R/2GM$), is such that
it is much easier to perform reliable time resolved spectroscopy for the brightest AGN,
where a typical flaring event may last hours or days.

There is in fact growing observational evidence for a number of Seyfert 1s,
that during a flare the X-ray spectrum becomes softer as the 2-10 keV flux
increases (see e.g. Done et al. (2000); Zdziarski \& Grandi (2001); 
Petrucci at al. (2001a); Vaughan \& Edelson \shortcite{ve01} 
and Nandra \shortcite{nan01} for a recent review).
More recently, Georgantopoulos \& Papadakis 
\shortcite{gp01} found
similar trends in a number of Seyfert 2 galaxies observed with {\it RXTE}. 

The analysis of the lightcurves in the frequency domain is usually presented in terms of
 Power Density Spectra (PDS). PDS of AGN are in general characterized by `red noise' 
variability
($P(f)\propto f^{-\gamma}$, with $ 1 \la \gamma \la 2$)
and a break at low frequencies \cite{lp93,me01}.
Red noise is characteristic of many astrophysical systems
(solar flares \cite{lh91}, Soft Gamma Repeaters bursts \cite{gog99}, 
Gamma-ray bursts \cite{bel00,mcb94}, etc.)
 as well as
more familiar ones (tides, earthquakes and even radio broadcasting, see e.g. Press 1978; Montroll
\& Shlesinger 1982), 
and is intimately related to the 
turbulent and/or chaotic nature of such non-linear systems. 
Nevertheless, the information contained in
the PDS alone is not enough to discriminate
between different mechanisms that many authors have proposed as the
origin of the $1/f$ noise.

In one of such attempts, Mineshige et al. \shortcite{mine94} (see also Mineshige \& Negoro 
1999)
 have proposed that black holes accretion flows develop into a state of
Self-Organized Criticality (Bak, Tang \& Wiesenfeld 1987) in which
mass accretion occurs in avalanches, whose size is distributed as a power-law: large
(and longer) shots contribute to the low-frequency part of the PDS, while small and shorter
shots determine the power-law decline at high frequencies.

In general, in order to explain the observed values of $\gamma$ in AGN and GBHC, any model 
requires individual signals of very
different timescales. 
In this way Poutanen \& Fabian \shortcite{pf99} were able to
reproduce the properties of the observed Power Spectrum of the GBHC Cygnus X-1 using
a simple stochastic pulse-avalanche model, originally proposed by Stern and  Svensson
\shortcite{ss96} to explain $\gamma$-ray burst temporal characteristics. In such
model the individual signals correspond to coronal hard X-ray magnetic flares, which possess
 a range of
durations and are capable to trigger larger avalanches.

Magnetic field reconnection of flux tubes in a tenuous corona is in principle
capable of producing power-law distribution of active region sizes. In pioneering work,
Tout \& Pringle \shortcite{tp96} have demonstrated both with a simple analytic model and
with Monte Carlo simulation, that, if  a dynamo-generated
magnetic field with a coherence scale of the order of the disc thickness emerges from the accretion disc, 
then the inverse cascade process of  stochastic magnetic reconnection of flux tubes may
lead to  much larger coherent fields and active regions, with a power-law distribution
in size.

Recently, Uttley \& McHardy (2001) have discovered a remarkable linear correlation between
RMS variability and flux in two X-ray binaries (Cygnus X-1 and the millisecond pulsar SAX J1808.4-3658).
They have also found that the lightcurves of three Seyfert galaxies are 
consistent with such relationship and have therefore argued that such behaviour is an universal one
in compact accreting systems.

Here we propose a simple model of Seyfert 1s high energy emission 
that combines the basic properties
the accretion flows must have to reproduce the short timescales temporal {\it and spectral}
characteristics of the observed objects. 

The basic idea of our model is that magnetic reconnection in the accretion disc corona does not
occur randomly in time and space, but rather in avalanches of correlated events (Poutanen \& Fabian 1999).
This not only affects the properties of the lightcurve, but, we argue, is responsible for the
observed spectral variability. The spectrum produced in region where 
a large number of heating events take place is different from the spectrum generated by
an isolated event, as we discuss in the next section. Furthermore, large avalanches of
magnetic flares may obscure the underlying disc as seen from the observer, and complicates
the temporal behaviour of the secondary spectral features (reflection hump and fluorescent lines). 

To build a full time-dependent model on the basis of the above ideas, 
including a detailed treatment of the radiative transfer in the accretion disc -- corona system
is undoubtedly a formidable task. Here we have chosen to simplify it by adopting an approximate
analytic treatment for the spectral generation, following the works of 
Di Matteo, Celotti \& Fabian (1997; 1999) and of Wardzi\'nski \& Zdziarski (2000) 
(as discussed in Appendix A). This, we believe, gives a fair representation of the 
modeled system and enables us to highlight the main prediction of the model in the simplest possible 
way.

The paper is structured as follows: in the next section we outline
 the main features of our model, 
dealing separately with its spectral (section \ref{structure}) and temporal (section \ref{time})
structure, and we elaborate on the thundercloud analogy (section \ref{thunder}). The results of 
numerical simulation and a discussion
of the main physical consequences of the model are presented in sections \ref{results}
and \ref{discussion}, respectively. Finally, in section \ref{conclusion} we  
draw our conclusions.

\section{Model outline}
\label{model}
\subsection{Structured coronae and thermal Comptonization}
\label{structure}
The best model to date that explains the spectra of Seyfert 1 galaxies is based on  thermal
Comptonization of soft photons in a hot electron cloud
(Shapiro, Lightman \& Eardley 1976; Sunyaev \& Titarchuk 1980).
Although it was early shown that the main features of the high-energy emission
of black hole powered sources could be explained by the interplay of the hot cloud 
(the so-called {\it corona})
shining on a standard (optically thick, geometrically thin) accretion disc
(Haardt \& Maraschi 1991), the geometry of the hot and
cold  phases is still matter of debate.
Such geometry could in principle be deduced with sufficient spectral information:
for example,
reflection features produced by a cold, optically thick medium illuminated by the
hard X-ray power-law continuum are sensitive to the detail of the relative
geometry of the two components \cite{geo91,mat91,mag95}. 
Simultaneous
studies of UV (likely produced by thermal emission from the optically thick medium)
and X-ray radiation have already been used in many cases to rule out a slab corona in favour of
a patchy one, made of a number of separate active regions
\cite{hmg94,ste95,dcf99}, while the observed correlation between the 
photon index and the reflection strength \cite{zls99,gil00}
(which has been proved to hold on timescales much longer than the dynamical one)
has demonstrated the need for a further geometrical/dynamical parameter, such as
the relativistic bulk motion velocity of the coronal material (Beloborodov 1999a; Malzac et al. 2001) 
and/or the truncation radius of the inner accretion disc \cite{zls99}.

As we focus here on  short-term variability, we do not consider these latter possibilities
(namely, we consider a {\it static} corona and a disc extending down to the innermost
stable orbit) and 
assume that the X-ray spectrum is produced by thermal
Comptonization in spherical active regions of size $R=rR_{\rm
S}$, lifted above the disc to a height $H=hR_{\rm S}$ (Di Matteo,
Celotti \& Fabian 1999), and permeated by a magnetic field with
intensity $B$. The seed photons for Comptonization are both
synchrotron ones, produced locally by the interaction
between the hot electrons and the magnetic field, and black-body
ones coming from the underlying cold disc.

 An active region illuminates the cold disc: part of the flux
is reflected and part is absorbed and reprocessed. The geometry
of the active region plays a crucial role in determining the
comptonized spectrum: if $L(r)$ is the 
luminosity of an active region, we have for the reprocessed luminosity
that returns to the active region and is
comptonized there (Beloborodov 1999; Malzac, Beloborodov \&
Poutanen 2001)
\begin{equation}
\label{lrep}
L_{\rm rep}(r) = L(r) (1-a) (1-\mu_0)/2,
\end{equation}
 where $a$ is the disc
albedo (that will be kept fixed to the value of 0.15 throughout the paper), 
and $\mu_0$ is a geometric factor which regulates the
feedback mechanism. 
As in Di Matteo, Celotti \& Fabian (1999), we
define a circular area of radius $R_{\rm hot}=r_{\rm hot} R_{\rm S}$ in the accretion
disc heated by an active region up to a temperature that depends on
the illuminating flux, and define
\begin{equation}
\mu_0=h/(r_{\rm hot}^2+h^2)^{1/2}
\end{equation}
 as the angular size of that circle measured from the center
of the active region (at height $h$). As a typical relevant
radius we take that one at which the temperature has decreased by
a factor of 2, and we have $r_{\rm hot} \simeq 2.3 h+r$.

If most of the accretion power is dissipated in the corona, and
the reprocessed radiation is the dominant source of soft photons
for Comptonization, the ratio $h/r$ determines the amplification
factor, $A(r)$, and the active region temperature ($\Theta=kT_{\rm e}/m_{\rm
e} c^2$): $A(r)=L/L_{\rm rep}=1+4\Theta+16\Theta^2$. The
resulting photon spectral index, $\Gamma$, can be related to $A$
by the approximate  expression (Beloborodov 1999a,b) $\Gamma \approx
2.33(A-1)^{-\delta}$, where $\delta\approx 1/6$ for GBHC and
$\delta\approx 1/10$ for AGNs. Clearly, for very large active
regions ($h/r \rightarrow 0$), the amplification
factor  has a minimum and the spectrum is softer.

In the more general situation the dominant source of soft photons
can be determined by comparing the local energy densities of the
synchrotron radiation and of the thermal emission
emerging from the disc, both from intrinsic dissipation and
reprocessed radiation.
As already discussed by Di Matteo, Celotti \& Fabian (1997, 1999)
and Wardzi\'nski \& Zdziarski (2000), the relevance of synchrotron
emission in a magnetically dominated active region decreases with
increasing central source mass (so it is less important in
AGN) and is a strong function of the temperature (the hotter the
corona the more it is important). Therefore, in the coronal flow around an active galactic
nucleus, synchrotron radiation is likely to be
relevant only for very compact active regions (see Appendix A).

For each active region size $r$ we calculate the
spectrum self-consistently: given an active region luminosity $L(r)$,
we assume that the energy density of the magnetic field is in
(almost) equipartition with that of the local radiation (Di
Matteo, Celotti \& Fabian 1997):
\begin{equation}
\frac{B^2}{8 \pi}=\frac{3 \epsilon_{\rm M} L(r) t_0}{4 \pi (r
R_{\rm S})^3},
\end{equation}
where $\epsilon_{\rm M}$ is the equipartition factor, of the order
of unity, and $t_0(R)=R/v_{\rm diss}$ is the dissipation time. The
dissipation velocity $v_{\rm diss}$ depends on the uncertain
nature of the heating process. For magnetic reconnection models
it is related to the Alfv\'en speed $v_{\rm A}$ and to the
magnetic Reynolds number $R_{\rm m}$ (which is very large in the
low resistivity coronal plasma) and ranges from $v_{\rm
A}/\sqrt{R_{\rm m}}$, in Sweet-Parker reconnection models \cite{swe58,par79}, 
to $v_{\rm A}/\ln{R_{\rm m}}$, in Petschek reconnection models 
\cite{pet64}. In the following (see
section \ref{time}), on the basis of time variability argument, we
will show that $v_{\rm diss}$ cannot be larger than a few percent
of the speed of light, and discuss further how to constrain the dissipation time.

We calculate the self-absorbed synchrotron luminosity
using the expression given in Wardzi\'nski \& Zdziarski (2000);
the intrinsic disc luminosity, instead,  
is simply given by $L_{\rm int}=\dot m (1-f_{\rm H}) L_{\rm Edd}$, 
where $\dot m$ is the accretion rate in units of the Eddington one, $ f_{\rm H}$ is
the fraction of the accretion power dissipated in the corona
(Svensson \& Zdziarski 1994) and $L_{\rm Edd} = 1.3 \times 10^{38} M/M_{\odot}$
is the Eddington luminosity.
Finally, the reprocessed luminosity, $L_{\rm
rep}(r)$, is calculated from Eq. (\ref{lrep}). The soft photon input flux from
the accretion disc is then given by 
the sum of the two thermal components at different temperatures, the 
intrinsic and the reprocessed one (see Appendix A for details).

In the following calculations of our fiducial model we will fix the overall
normalization constants $M_{\rm BH}=10^7 M_{\odot}$,
$\dot m=0.1$ and $f_{\rm H}=0.7$. Furthermore, we neglect radial
dependence of the energy output and consider only the coronal flow above the inner, more
luminous part of the accretion disc, so that $R_{\rm disc} = 20 R_{\rm S}$.

To calculate the comptonized spectrum we use the analytic
expression given in Wardzi\'nski \& Zdziarski (2000) for a
homogeneous and isotropic source characterized by optical depth
$\tau_{\rm T}$ and temperature $\Theta$. To derive the emerging spectrum 
we fix the coronal optical depth, $\tau_{\rm T}$, the active region size $r$ and its
luminosity $L(r)$ (see section \ref{thunder}), and
calculate the temperature by imposing that the total luminosity of the active region
(synchrotron plus inverse Compton emission) equals $L(r)$ (as discussed in Appendix A).

In Fig. 1 we plot the coronal temperature, $\Theta$, and the slope of the
comptonized continuum, $\alpha$, emerging from each active region 
as functions of an active region size, for different values of
the optical depth. As anticipated, larger and  more luminous regions are cooler,
and henceforth  softer,
 due to the larger reprocessed soft radiation they intercept.
Although calculated for a spherical active region, for the sake of consistency 
we have checked that, in the limits
$h/r \rightarrow \infty$ and $f_{\rm H}\rightarrow 1$, our results agree with the prediction
of Haardt \& Maraschi (1993) for a slab corona in energy equilibrium above a cold accretion 
disc. 

\begin{figure}
%\vbox to115mm{\vfil
\psfig{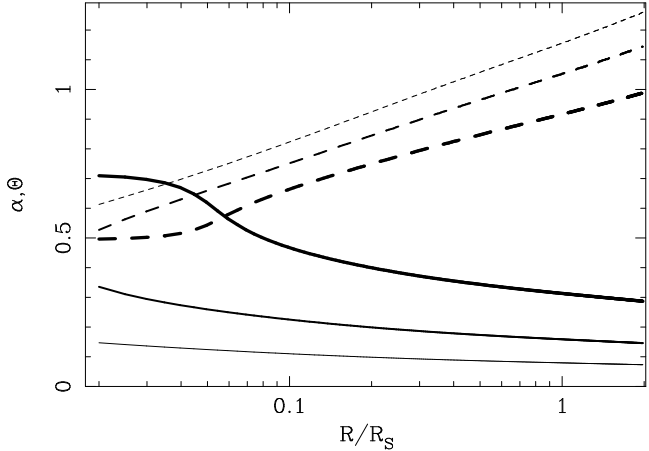}
\caption{The equilibrium temperature ($\Theta=kT/m_{\rm e}c^2$, solid lines) 
and spectral index $\alpha=\Gamma-1$ (dashed lines)
of an active region as a function of its size, for a fixed height
$H=1R_{\rm S}$ and different optical depths, calculated for $C=10^{-3}$ and $D=1$ (see section \ref{thunder}). 
In order of decreasing line-thickness, the different curves 
correspond to $\tau_{\rm T}=0.5,1,2$. Due to the increased
reprocessed radiation that is intercepted in the larger regions,
the temperature drops with increasing $r$, and the emerging
spectrum is softer. The shoulder at low values of $R$ for the $\tau_{\rm T}=0.5$ case is due to the 
increased importance of synchrotron emission as a source of soft photons (see also Figure A1).} 
%\vfil} 
\label{theta}
\end{figure}

\subsection{Active regions distribution and luminosity: the thundercloud analogy}
\label{thunder}
In the hot, comptonizing corona the
energy input is likely to be provided by a strong, highly intermittent magnetic field
(see Merloni \& Fabian 2001) amplified in the underlying turbulent disc by a dynamo
process generated by the Magneto Rotational Instability (Balbus \& Hawley 1998;
Tout \& Pringle 1992). Dissipation is then triggered by magnetic reconnection of
flux tubes on the smallest possible scale (given approximately by the disk scaleheight).
If these small reconnection sites were distributed randomly in time and space, all
with a similar aspect ratio, $h/r$, we would
 observe shot-noise-like variability in the lightcurve and, more importantly, 
 no spectral variability.
If, on the other hand, the system is in some kind of critical
state, such that each small flare can trigger an avalanche in its
immediate neighborhood (i.e. the flares are spatially and temporally correlated), the situation
will be qualitatively different. In fact, any
such avalanche is effectively a large active region: more
luminous (because it contains more reconnection sites), more
extended and with a softer spectrum (because $h/r$ is smaller, see previous section).

As a consequence of such avalanche generation process at any given time 
there would be a spread of active regions above the disc,
with typical size ranging from the very small micro-flares ($R \sim$ disc scaleheight),
to the larger avalanches ($R \sim$ a few Schwarzschild radii).

Hence, we assume that the luminosity of an active region is determined
by the size of the avalanche, and therefore scale as the size of the region itself
\begin{equation}
L(r)=c_1 \dot m f_{\rm H} L_{\rm Edd} r^D.
\end{equation}
Here $D$ may be related to the spatial distribution of the
correlated flares (and be regarded as a fractal dimension of the
reconnection sites) and/or to the radial dependence of the energy
generation in the accretion disc.

On the other hand, because of the stochastic nature of the 
avalanche generation process, we will assume a {\it power-law} shape for the
 time-independent active region size distribution $n(r)$ 
(as also suggested by the shape of the observed power density spectra), such that the  number
of active regions of size $r$ lying in the interval $(r,r+dr)$ is, {\it at any time}, given by
\begin{equation}
n(r)dr=c_2 r^{-p} dr.
\end{equation}
The two constants $c_1$ and $c_2$ can be fixed imposing the overall normalization
for the total corona average covering fraction $C$ and for the average hard luminosity:
\begin{eqnarray}
C&=&\int_{r_{\rm min}}^{r_{\rm max}} n(r) \frac{{\cal A}(r)}{A_{\rm disc}} dr \\
L_{\rm hard}&=&\dot m  f_{\rm H} L_{\rm Edd} =
\int_{r_{\rm min}}^{r_{\rm max}} n(r) L(r) dr ,
\end{eqnarray}
where ${\cal A}(r)$ is the area of the disc covered by an active region
of size $r$ and $A_{\rm disc}$ is the total area of the disc.

To summarize, we model the accretion disc corona as highly inhomogeneous
stochastic system, whose basic building blocks, the active regions, can be viewed
as `magnetic thunderclouds', charged by the differential rotation of the underlying disc
and/or the turbulent motions in the accretion flow. The sizes of the thunderclouds
are distributed as a power-law. The fast energy release, triggered
by magnetic reconnection on the smallest scales, heats progressively larger active regions.
Each active region (thundercloud) produces the observed rapid flares (X-ray lightning strokes) 
by inverse Compton scattering soft photons
coming mainly from the underlying optically thick accretion disc\footnote{It is interesting here to
remark that the analogy between high energy astrophysical processes and terrestrial lightning is not new.
McBreen et al. (1994) have pointed out the similarity between the log-normal properties of gamma-ray bursts
lightcurves and those of terrestrial lightning. In the latter case, the durations, peak currents, intervals
between the strokes in the flashes, and the flash charges are log-normally distributed (Berger et al. 1975).
When the dispersion of a log-normal distribution is large, the distribution is mimicked by a 
$1/x$ distribution over a wide range of $x$, while it can be shown that amplification processes generally 
relate log-normally distributed variables 
to power-law tails (Montroll \& Shlesinger 1982).}. 
Furthermore,
if the coronal optical depth $\tau_{\rm T}$ is high enough, the active regions may 
obscure the X-ray spectral features produced 
 in the cold disc (in particular the fluorescent lines and the reflection hump, as discussed in
 Matt, Fabian \& Reynolds 1997 and
Petrucci et al. 2001b, respectively) 
for an observer situated above them (and act therefore as `Compton clouds'). 

\subsection{Time variability and Power Density Spectra}
\label{time}

In general, if the heating process is related to the dissipation
of the magnetic field in the corona, the
dissipation timescale is given by $t_0(R) \simeq 10^2 r M_7 b$ s, where $M_7=M_{\rm
BH}/10^7 M_{\odot}$ and $b=c/v_{\rm dis}$ (Haardt, Maraschi \&
Ghisellini 1994). Here we assume that the spectrum is produced instantaneously in
every active region, i.e. the timescale over which
an active region is heated is longer than the region
light-crossing time and $b \ga$ a few. We fix $b=30$ (see section \ref{mag} for a 
discussion of this point).

Power density spectra
of both AGNs and GBHCs show a break that separates low-frequency white noise
($P(f)\sim$ constant) from the red noise part of the PDS. Within our model, 
such a break can naturally
be associated with  the largest possible active region, and thus fix $r_{\rm max}$.
On the other hand, in any stochastic process in which the characteristic durations
and amplitudes are distributed as power-laws, we have
\begin{equation}
N(t_0)I^2(t_0)\propto t_0^\beta \;\; {\rm for}\;\; t_{0,{\rm min}}<t_0<t_{0,{\rm max}}
\end{equation}
where $N(t_0)$ is the probability and $I^2(t_0)$
the mean square amplitude of a flare with a duration $t_0$ and the PDS is 
approximately given by
\begin{equation}
P(f)\propto f^{-\gamma} \;\;\;\; \gamma=(3+\beta),
\end{equation}
in the range $(2\pi t_{0,{\rm max}})<f<(2\pi t_{0,{\rm min}})$
(Lochner, Swank \& Szymkowiak 1991). Then, according to our
model, for which $N(t_0)\propto t_0^{-p}$ and $I(t_0)\propto t_0^D$, the
observed slope of the PDS determines the relation between $p$ and
$D$:
\begin{equation}
\label{alpha}
p=2D+3-\gamma.
\end{equation}
In the following section we present a numerical simulation of
the thundercloud model for AGN spectral/temporal variability, keeping fixed
$\gamma=1.5$ (as usually dictated by the observations of AGN), and 
we discuss the main observable consequences of the thundercloud model.

\section{The full picture: results from simulated lightcurves}
\label{results}
In this section we describe the results we obtained 
from simulated lightcurves with variable X-ray 
spectra.
To take into account time dependence we introduce a new distribution $n(r,t)$, defined in such a way
that the number of active regions of size in the interval $(r,r+dr)$ {\it generated} in the time bin
$(t,t+dt)$ is given by $n(r,t)\,dr\,dt$. In a stationary state $n(r,t)$ does not depend on
$t$, and we assume $n(r)=n(r,t) t_0(r)$, where $t_0(r)$ is the total duration of an avalanche. 

\begin{figure}
%\vbox to120mm{\vfil
\psfig{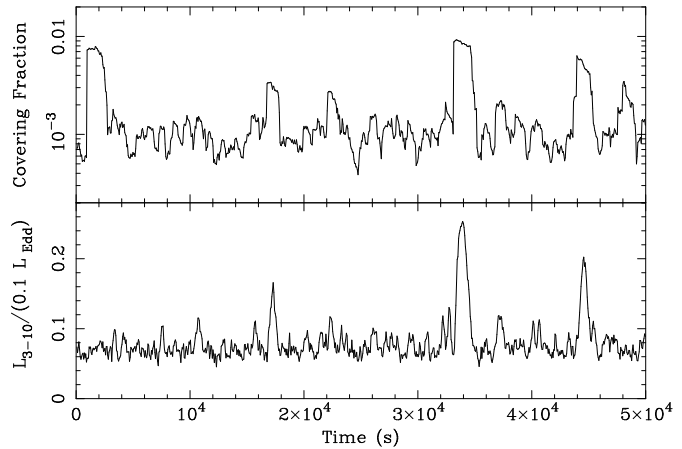}
\caption{A simulated lightcurve (bottom panel) with its corresponding evolution of the 
covering fraction 
(top panel), for a $M_{\rm BH}=10^7 M_{\odot}$ central black
hole accreting at one tenth of the Eddington rate.} 
\label{lc_cove}
\end{figure}

\begin{figure}
\vbox to120mm{\vfil
\psfig{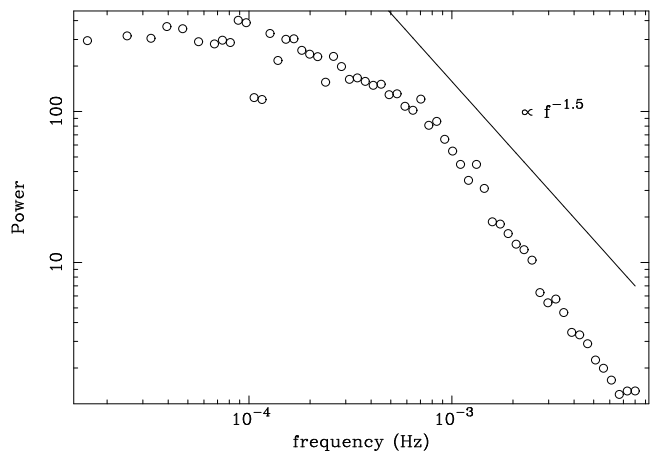}
\caption{Power density spectrum of the  simulated lightcurve shown in Fig. \ref{lc_cove}.
The indexes of
the active region distributions are $D=1$ and $p=3.5$. To reduce the scatter the 
PDS has been obtained averaging the power density spectra of 16 lightcurves of the total duration of 
$1.2\times 10^5$ seconds each.} 
\vfil}
\label{fig_pds}
\end{figure}

A simulated lightcurve (see Fig. \ref{lc_cove} for an example) is then produced
requiring that in any time bin the actual number $\cal N$$(r,t)$
of avalanches generated with   
 size in the interval $(r,r+dr)$ be  picked from a
Poissonian distribution with mean $n(r,t)dr$ and then summing over all the active 
regions generated at any time. 
For the sake of simplicity we 
adopted a triangular shape for the individual flares profile. 
Because we enforce relation (\ref{alpha}) for the indexes of the distribution, the final PDS
has the correct slope in the red-noise part (see Figure \ref{fig_pds}).

The emerging
spectrum is calculated superimposing the contribution from all
the flaring regions active at that time. The photon index
$\Gamma$ is calculated for every time bin with a least square fit
to the spectrum with a power-law model in the 3-10 keV band.

We have fixed $r_{\rm min}=0.02$ and $r_{\rm max}=4$. The average covering
factor $C$, in turn,  fixes the total number of active regions and the
degree of variability (RMS) of the PDS.

To explore the parameter space, we have simulated AGN light curves 
of total duration equal to $2.4\times 10^6$ s, with a temporal resolution
of $60$ s. We  varied the coronal optical depth (in the range
$0.5 \le \tau_{\rm T} \le 2$), the average covering fraction ($5\times 10^{-4} \le C \le 2\times 10^{-3}$)
and the luminosity scaling index ($0.6\le D \le 1.4$). 

\subsection{The spectral index-luminosity correlation}
\begin{figure}
%\vbox to110mm{\vfil
\psfig{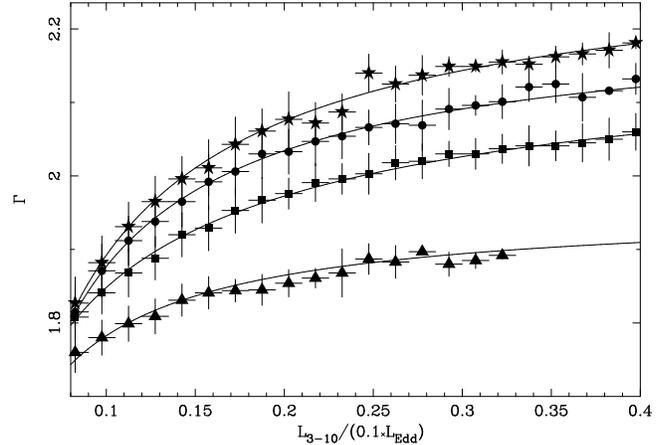}
\caption{The observed photon index in the 3-10 keV band plotted versus 
the luminosity in the same band (in units of $0.1 L_{\rm Edd}$) as obtained from our fiducial simulation for $D=1$
and $C=10^{-3}$. 
The points for all the time intervals have been grouped in luminosity
bins and the average gamma has been calculated. The error bars correspond to the weighted statistical 
uncertainty. Also shown are the best fitting functions (as in Table 1) for each of the four groups.
Triangles corresponds to $\tau_{\rm T}=0.5$, squares to  $\tau_{\rm T}=1$,
circles to  $\tau_{\rm T}=1.5$ and stars to $\tau_{\rm T}=2$.} 
%\vfil} 
\label{L_gamma}
\end{figure}

In Figure \ref{L_gamma}, for $D=1$ and $C=10^{-3}$, we plot the photon index
$\Gamma$ of the total emerging spectrum, measured in the 3-10 keV band, as a function of the
luminosity in the same band, for three different values of
the optical depth. 
As discussed in section \ref{model}, the higher the 3-10 keV luminosity, the softer 
the spectrum in that energy band, as observed. Furthermore, when the largest
flares occur, and consequently the instantaneous covering fraction has a maximum (see Fig. 
\ref{lc_cove}), the $\Gamma - L$ relation tends to saturate to the softest value corresponding 
to an almost slab-like geometry. 
This is an intrinsic property of our model, and is due to the fact that larger active regions are cooler
because the reprocessed soft luminosity that reenters the blobs (and is comptonized there) is larger.
 
\begin{figure}
%\vbox to110mm{\vfil
\psfig{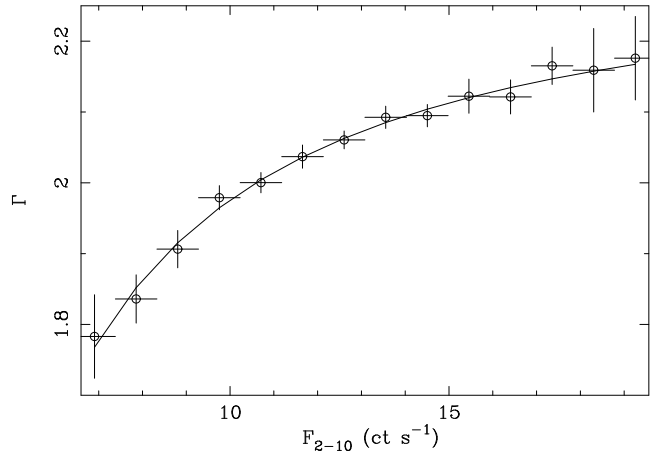}
\caption{The observed photon index in the 2-10 keV band plotted versus the RXTE count rate
in the same band for the Seyfert 1 galaxy MCG--6-30-15, from the analysis of Vaughan \& Edelson (2001) (their Figure 3). 
The data point have been grouped in flux bins and the errors correspond to the weighted statistical uncertainty 
on the value of the spectral index. Also shown is the best fitting function of the form $\Gamma=\Gamma_0-K F^{\delta}$.
We obtain $\Gamma_0=2.30^{+0.63}_{-0.02}$, $K=6.8^{+20.8}_{-4.5}$ and $\delta=-1.31^{+0.79}_{-0.79}$.} 
%\vfil} 
\label{gamma_f_sv}
\end{figure}

Overall, we find a spectral variability range
$\Delta \Gamma \sim 0.3-0.4$ for a change in luminosity of about a factor of four, 
similar to what is observed in many Seyfert 1s. 
It is worth remarking that a similar correlation was indeed predicted by 
 Haardt, Maraschi \& Ghisellini (1997) for the case of a pair dominated corona, but in that 
case the same $\Delta \Gamma$ was achieved for a luminosity variation of at least two orders 
of magnitude, which is clearly inconsistent with the observations. To verify that, we show in Figure 
\ref{gamma_f_sv} the observed $\Gamma - L$ correlation for the Seyfert 1 galaxy MCG--6-30-15 
(from Vaughan \& Edelson 2001) observed with {\it RXTE} in 1997\footnote{A thorough comparison of the 
thundercloud model
with ASCA data for MCG--6-30-15 will be presented elsewhere (Shih et al. 2001).}. 

For both our simulated lightcurves and the MCG--6 data we have 
fitted the $\Gamma$-luminosity relation with the function 
\begin{equation}
\Gamma = \Gamma_0-K L^{\delta}.
\end{equation}
 In Table 1 the values  of the three fitting parameters $\Gamma_0$, $K$ and  $\delta$ are listed for our set
of simulated lightcurves.
The asymptotic value of the photon index, $\Gamma_0$,
 depends mainly on the optical depth, and is larger for larger values of 
$\tau_{\rm T}$. The exponent $\delta$, instead, which determines the amount of spectral variation
for a given increase in luminosity, is mainly dependent on the index $D$. 

For the observed data plotted in Figure \ref{gamma_f_sv} we obtain 
$\Gamma_0=2.30^{+0.63}_{-0.02}$, $K=6.8^{+20.8}_{-4.5}$ and $\delta=-1.31^{+0.79}_{-0.79}$. 
Compared to our model simulations, such a result is a strong indication of a relatively high value for the
coronal optical depth ($\tau_{\rm T}\ga 1.5$) in this source, at least for the observation considered here.
This, in turn, agrees well with the results presented in Guainazzi et al. (1999) of the analysis of a 
Beppo{\it SAX} 
observation of the same source: the measured spectral cut-off energy 
($E_{\rm c} \simeq 130 \pm 40$ keV for their best-fitting model), in the framework of thermal Comptonization models
implies a coronal temperature\footnote{We note that, usually, an e-folding power-law is not a good approximation
for the actual shape of the observed high energy cut-off in many sources, in particular those with the better
quality data, see e.g. Frontera et al. (2001). The actual coronal temperature is somewhat model and geometry 
 dependent, and its real value for 
MCG--6-30-15 may differ from the one we adopt here by a factor of the order of unity, which does not change
our main conclusion regarding the source optical depth.}
$\Theta \sim 0.1$, 
that we obtain for $\tau_{\rm T}\ga 1.5$ (see Fig. \ref{theta}).  
Although it is more difficult to put tight constraints on the other two model parameters ($D$ and $C$), nonetheless 
we may conclude that the 
observed $\Gamma - L$ correlation is consistent with $D\sim 1$ and $C \sim 10^{-3}$.
 Such a fairly small value for the average covering fraction is indeed 
required by the observed data: the smaller the covering
fraction, the larger the observer variability, and the greater the chance of a large flare to occur. 
Too a large value of $C$ would therefore drastically reduce the observed spectral variability.
For our fiducial case, $D=1$ and $C=10^{-3}$, we have $c_1=0.16$ and $c_2=0.046$.
The total number of active regions at any time is on
average\footnote{The total number of individual micro-flares
that are the fundamental constituents of an avalanche is of
course larger, depending on the actual number of events that are
contained in any larger active region.} $\cal N$$=\int n(r) dr \simeq 160$.

\begin{table}
% \centering
% \begin{minipage}{140mm}
  \caption{Results from simulations and best fit parameters for the $\Gamma$-luminosity relation. We fitted 
a function of the form $\Gamma=\Gamma_0 - K L^{\delta}$.}
  \begin{tabular}{@{}cccccc@{}}
   \multicolumn{3}{c}{Model parameters} &  \multicolumn{3}{c}{Fit parameters}\\
   C & D  & $\tau_{\rm T}$ & $\Gamma_0$ & K & $\delta$ \\ [2pt]
\hline \\
$0.0005$ & $0.6$ & $0.5$ & $2.01^{+0.04}_{-0.02}$ & $0.050^{+0.041}_{-0.019}$ & $-0.55^{+0.20}_{-0.20}$ \\[5pt]
$0.0005$ & $0.6$ & $1.0$ & $2.21^{+0.11}_{-0.04}$ & $0.08^{+0.12}_{-0.03}$    & $-0.49^{+0.23}_{-0.22}$ \\[5pt]
$0.0005$ & $0.6$ & $1.5$ & $2.30^{+0.01}_{-0.02}$ & $0.058^{+0.019}_{-0.013}$ & $-0.65^{+0.11}_{-0.11}$ \\[5pt]
$0.0005$ & $0.6$ & $2.0$ & $2.34^{+0.01}_{-0.01}$ & $0.030^{+0.011}_{-0.008}$ & $-0.93^{+0.13}_{-0.14}$ \\[5pt]
$0.0005$  & $1.0$ & $0.5$ & $2.07^{+0.01}_{-0.02}$ & $0.11^{+0.45}_{-0.07}$    & $-0.37^{+0.23}_{-0.41}$ \\[5pt]
$0.0005$  & $1.0$ & $1.0$ & $2.21^{+0.18}_{-0.06}$ & $0.10^{+0.25}_{-0.05}$    & $-0.56^{+0.26}_{-0.26}$ \\[5pt]
$0.0005$  & $1.0$ & $1.5$ & $2.32^{+0.13}_{-0.05}$ & $0.10^{+0.01}_{-0.003}$   & $-0.62^{+0.22}_{-0.22}$ \\[5pt]
$0.0005$  & $1.0$ & $2.0$ & $2.32^{+0.01}_{-0.01}$ & $0.040^{+0.013}_{-0.009}$ & $-1.00^{+0.11}_{-0.12}$ \\[5pt]
$0.0005$  & $1.4$ & $0.5$ & $1.96^{+0.01}_{-0.01}$ & $0.019^{+0.072}_{-0.009}$ & $-0.89^{+0.68}_{-0.92}$ \\[5pt]
$0.0005$  & $1.4$ & $1.0$ & $2.09^{+0.01}_{-0.01}$ & $0.010^{+0.012}_{-0.009}$ & $-1.28^{+0.63}_{-0.64}$ \\[5pt]
$0.0005$  & $1.4$ & $1.5$ & $2.15^{+0.35}_{-0.05}$ & $0.085^{+0.43}_{-0.005}$  & $-1.44^{+1.22}_{-1.70}$ \\[5pt]
$0.0005$  & $1.4$ & $2.0$ & $2.38^{+0.02}_{-0.02}$ & $0.070^{+0.01}_{-0.003}$  & $-0.81^{+0.50}_{-0.79}$ \\[5pt]
$0.001$ & $0.6$ & $0.5$ & $2.09^{+0.31}_{-0.07}$ & $0.13^{+0.30}_{-0.05}$       & $-0.35^{+0.22}_{-0.22}$ \\[5pt]
$0.001$ & $0.6$ & $1.0$ & $2.20^{+0.10}_{-0.05}$ & $0.075^{+0.096}_{-0.035}$    & $-0.62^{+0.24}_{-0.24}$ \\[5pt]
$0.001$ & $0.6$ & $1.5$ & $2.36^{+0.05}_{-0.06}$ & $0.13^{+0.05}_{-0.005}$      & $-0.46^{+0.10}_{-0.19}$ \\[5pt]
$0.001$ & $0.6$ & $2.0$ & $2.36^{+0.05}_{-0.02}$ & $0.081^{+0.052}_{-0.030}$    & $-0.63^{+0.15}_{-0.16}$ \\[5pt]
$0.001$  & $1.0$ & $0.5$ & $1.96^{+0.02}_{-0.01}$ & $0.024^{+0.003}_{-0.005}$    & $-0.88^{+0.77}_{-0.76}$ \\[5pt]
$0.001$  & $1.0$ & $1.0$ & $2.21^{+0.34}_{-0.05}$ & $0.11^{+0.06}_{-0.02}$    & $-0.54^{+0.30}_{-0.31}$ \\[5pt]
$0.001$  & $1.0$ & $1.5$ & $2.23^{+0.02}_{-0.02}$ & $0.048^{+0.030}_{-0.022}$    & $-0.86^{+0.49}_{-0.48}$ \\[5pt]
$0.001$  & $1.0$ & $2.0$ & $2.32^{+0.14}_{-0.08}$ & $0.071^{+0.15}_{-0.046}$     & $-0.79^{+0.29}_{-0.30}$ \\[5pt]
$0.001$  & $1.4$ & $0.5$ & $2.04^{+0.08}_{-0.03}$ & $0.019^{+0.072}_{-0.009}$ & $-0.89^{+0.68}_{-0.92}$ \\[5pt]
$0.001$  & $1.4$ & $1.0$ & $2.17^{+0.01}_{-0.02}$ & $0.010^{+0.012}_{-0.009}$ & $-1.28^{+0.63}_{-0.64}$ \\[5pt]
$0.001$  & $1.4$ & $1.5$ & $2.22^{+0.09}_{-0.09}$ & $0.085^{+0.43}_{-0.005}$  & $-1.44^{+1.22}_{-1.70}$ \\[5pt]
$0.001$  & $1.4$ & $2.0$ & $2.34^{+1.43}_{-0.16}$ & $0.057^{+0.051}_{-1.196}$  & $-0.85^{+0.70}_{-0.72}$ \\[5pt]
$0.002$  & $0.6$ & $0.5$ & $1.92^{+0.02}_{-0.02}$ & $0.0011^{+0.001}_{-0.0005}$ & $-2.1^{+0.3}_{-0.4}$ \\[5pt]
$0.002$  & $0.6$ & $1.0$ & $2.17^{+0.14}_{-0.05}$ & $0.077^{+0.133}_{-0.042}$   & $-0.63^{+0.29}_{-0.29}$ \\[5pt]
$0.002$  & $0.6$ & $1.5$ & $2.32^{+0.01}_{-0.01}$ & $0.13^{+0.03}_{-0.01}$      & $-0.52^{+0.71}_{-0.67}$ \\[5pt]
$0.002$  & $0.6$ & $2.0$ & $2.39^{+0.14}_{-0.06}$ & $0.11^{+0.13}_{-0.04}$      & $-0.61^{+0.22}_{-0.22}$ \\[5pt]
$0.002$  & $1.0$ & $0.5$ & $1.96^{+0.07}_{-0.02}$ & $0.010^{+0.006}_{-0.003}$   & $-1.19^{+0.59}_{-0.62}$ \\[5pt]
$0.002$  & $1.0$ & $1.0$ & $2.11^{+0.03}_{-0.01}$ & $0.006^{+0.002}_{-0.001}$   & $-1.6^{+0.40}_{-0.40}$ \\[5pt]
$0.002$  & $1.0$ & $1.5$ & $2.17^{+0.11}_{-0.04}$ & $0.016^{+0.001}_{-0.0005}$  & $-1.26^{+0.58}_{-0.59}$ \\[5pt]
$0.002$  & $1.0$ & $2.0$ & $2.67^{+0.35}_{-0.14}$ & $0.29^{+0.35}_{-0.12}$      & $-0.43^{+0.19}_{-0.19}$ \\[5pt]
$0.002$  & $1.4$ & $0.5$ & $1.92^{+0.05}_{-0.03}$ & $0.015^{+0.035}_{-0.014}$  & $-1.21^{+0.68}_{-0.69}$ \\[5pt]
$0.002$  & $1.4$ & $1.0$ & $2.01^{+0.02}_{-0.01}$ & $0.002^{+0.001}_{-0.004}$  & $-1.75^{+0.76}_{-0.70}$ \\[5pt]
$0.002$  & $1.4$ & $1.5$ & $2.46^{+1.21}_{-0.16}$ & $0.21^{+0.92}_{-0.05}$     & $-0.43^{+0.32}_{-0.30}$ \\[5pt]
$0.002$  & $1.4$ & $2.0$ & $2.21^{+0.17}_{-0.06}$ & $0.025^{+0.096}_{-0.018}$  & $-1.04^{+0.45}_{-0.46}$ \\
\end{tabular}
%\end{minipage}
\end{table}

\subsection{The variability-luminosity correlation}
As already mentioned in the introduction, recent analysis of time variability properties of 
X-ray binaries and AGN (Uttley \& McHardy 2001) have highlighted an intrinsic property 
of broadband noise variability in accerting compact objects, namely the strong linear
correlation between RMS variability and flux.

\begin{figure}
%\vbox to100mm{\vfil
\psfig{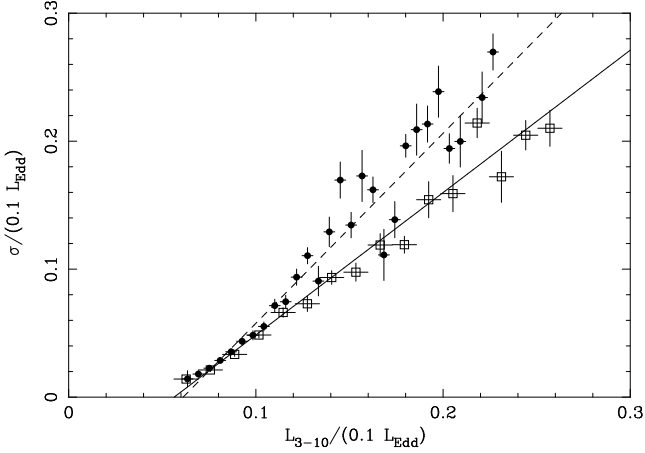}
\caption{Luminosity dependence of the variance calculted for segments of a simulated lightcurve 
($D=1$, $C=10^{-3}$
and $\tau_{\rm T}=1$) 
$7.7\times 10^3$ (filled circles, dashed best fitting line) and $1.9 \times 10^3$ 
(open squares, solid best fitting line) seconds long, with resolution of 60 seconds. 
The fitting functions are of the form $\sigma=k(L_{3-10}-L_0)$, with $k$ and $L_0$ constants.} 
%\vfil} 
\label{fig_uttley}
\end{figure}

We have analysed our simulated lightcurves to look for a similar correlation.
  With the parameters $D$ and $C$ fixed to the vaues of $1$ and $0.001$, respectively,
we have simulated a lightcurve of the total duration  of $4.3 \times 10^6$ s, subdivided
it in 560 segments of 7680 s each in the first case, and in 2240 segments of 1920 s 
each in the second case, and calculated the mean luminosity and the variance 
$\sigma$ for each segment in the standard way. 
The results, binnned in luminosity intervals, 
are shown in Fig. \ref{fig_uttley}. A clear correlation is evident in both cases,
consistent with a linear one. We fitted a function  of the form $\sigma=k(L_{3-10}-L_0)$, with $k$ and $L_0$ constants, and found
$k=1.49\pm0.05$ and $L_0=0.091\pm0.006$ for the long interval case (dashed line in Fig. \ref{fig_uttley}), and
$k=1.11\pm0.04$ and $L_0=0.063\pm0.005$ for the short interval case (solid line in Fig. \ref{fig_uttley}). 
In the framework of the thundercloud model such behaviour 
can be explained by the non-random nature of the flare distribution. 
In fact, the large number of small, hard, low luminosity active regions consitutes a  roughly constant 
background, with very low RMS variability, while  much higher variability is associated with
the rare large luminous events. The different slopes of the correlation in the two cases simply reflect
the fact that the length of the segments in which we subdivided the lightcurve is shorter than the
longest flare timescale, and we are sampling the PDS in its red-noise part.

\section{Discussion}
\label{discussion}
\subsection{Coronal scaleheight}
In the above calculation we have kept fixed the value of the
height of the flares above the accretion disc. In fact, as long
as the main source of soft photons for Comptonization is the reprocessed radiation from the 
disc itself (as in the case of our fiducial model for which 
$c_{\rm r} > c_{\rm i} \gg c_{s}$, see Appendix A), what matters for the
spectral properties of the source is the ratio $h/r$ of an active region height
over its size. Then, in order for the
model to accommodate spectral variations as large as those
observed in many Seyfert 1 galaxies, we need $r_{\rm
min}\ll h \ll r_{\rm max}$. 

If the small-scale flux loops are
generated by a self-sustaining magnetic dynamo operating inside
the disc, it is natural to identify the size of the smallest active
regions emerging from  the disc, i.e. the micro-flares that
trigger the larger avalanches, with the accretion disc thickness
\cite{tp96,grv79,nk01}.
Thus, the observed spectral variability implies that the
average elevation of a reconnection site is larger (and perhaps
much larger) than the disc scaleheight, as firstly proposed by Di
Matteo (1998). 

It is interesting that full MHD simulations of
magneto rotational instability in weakly magnetized accretion
discs indeed show the formation of a strongly magnetized corona,
considerably more extended in the vertical direction than the
disc itself (Miller \& Stone 2000). Moreover, in the framework of
a model similar to the one we present here, Di Matteo, Celotti \&
Fabian (1999) studied observations of the GBHC GX 339-4 in different
spectral states to place limits on the model parameters, and in
particular on the flare height. They conclude that the X-ray emission
from galactic black holes in their {\it hard} state is more
likely a result of inverse Compton scattering of soft photons in
hot magnetic flares triggered high above the accretion disc. 

The analytic treatment of Svensson \& Zdziarski (1994) of the accretion disc--corona system
allows us to give an estimate of the ratio of the coronal to disc scaleheight in the inner,
radiation pressure dominated, part of the disc. From such a calculation, we can conclude that
indeed, given the observed coronal temperatures, whenever most of the accretion power
is released in the corona (say, $f_{\rm H}\ga 0.7$),
the corona is much more extended in the vertical direction than the underlying disc. We have
\begin{eqnarray}
\frac{H_{\rm c}}{H_{\rm d}}&=&
\left(\frac{m_{\rm e}}{m_{\rm p}}\right)^{1/2} \frac{4}{9}
\varpi^{3/2} \Theta^{1/2} \dot m^{-1} J(\varpi)^{-1} (1-f_{\rm H})^{-1} \nonumber \\
\!&\ga & 16 \left(\frac{\Theta}{0.1}\right)^{1/2} \left(\frac{\dot m}{0.1}\right)^{-1}
\left(\frac{(1-f_{\rm H})}{0.1}\right)^{-1},
\end{eqnarray}
where $\varpi$ is the radial disc coordinate in units of Schwarzschild radii,
$J(\varpi)=1-\sqrt{3/\varpi}$, and the last inequality is obtained 
taking the minimum of $\varpi^{3/2}/J(\varpi)$.

In the case of a structured corona the situation ought to be more complicated:
if the flux tubes emerging from the disc carry a magnetic field with strength comparable
to the equipartition with the internal disc pressure, they will tend to expand (reducing
the strength of the magnetic field) in the less pressurized coronal atmosphere
(Parker, 1979, \S 8.4). However, if the flux tubes are twisted,
as is likely to be the case if they are anchored in the
differentially rotating disc underneath,
their sideways expansion is counterbalanced by the generation
inside the flux tube itself of an azimuthal field component, with a net increase of its total
magnetic energy content. The flux rope will rise buoyantly
(Parker, 1979, \S 9.1, 9.6), keeping a much lower aspect ratio (defined as the ratio of
the loop length over its diameter). To conclude, given the major uncertainties on the magnetic
flux tube dynamics in the disc-corona interface, 
we believe that it is not unlikely for
the height of the magnetic flares in the corona to be larger than the disc thickness\footnote{As a 
final remark, we note that a puzzling characteristic
of solar coronal loops, so often invoked in analogy with accretion disc coronal ones,
is their tendency to have a nearly uniform thickness, instead of being
substantially wider at their tops than at their foot-points, as expected from a
straightforward application of the force-free assumption
(Klimchuk, Antiochos \& Norton, 2000).}, as envisaged in our model, and, indeed, needed
to explain the large observed spectral variability.

\subsection{The magnetic field}
\label{mag}
Within the framework of active coronal models for the hard X-ray 
emission of black hole accretion flows, 
the exact nature of the heating process remains the most elusive issue.
Here, following the argument of Merloni \& Fabian (2001), we have assumed that the
magnetic field is the main repository of the energy in the corona and that the active regions
are heated by the dissipation of such energy, probably via magnetic reconnection. 

Although there is a growing consensus on the dynamical importance of magnetic fields in 
accretion discs, mainly driven by recent progresses in numerical MHD simulation
of accretion flows and magneto--rotational instability, the 
detailed topology of the magnetic field is 
far beyond the reach of our current theoretical understanding.  
The solar analogy has provided theorists with ideas and inspiration, if not with quantitative 
insight, but it is wise to bear in mind that although we are able to actually image magnetic structures
in our nearest star, even the heating of the solar corona is still not well understood.
 
With these {\it caveats} in mind, we discuss here some of the implications of our
stochastic model for the accretion disc corona on the strength of the magnetic field and on the
heating mechanism.

The ratio of magnetic to thermal energy in an active region is given by 
\begin{eqnarray}
E_{\rm mag}/E_{\rm th} & \! \simeq \! & 
\frac{\epsilon_{\rm M} c_1 r^D \dot m f_{\rm H} L_{\rm Edd} t_0 \sigma_{\rm T}}{4 \pi \tau_{\rm T} R^2 k T_{\rm e}} \nonumber \\ 
& \! \sim\! &  1.3 \times 10^3 \frac{\epsilon_{\rm M} \dot m f_{\rm H} b}{\tau_{\rm T} \Theta} c_1 r^{D-1}.
\end{eqnarray}
For the standard set of physical parameters we have adopted, 
as long as the energy in the magnetic field is more than a few percent of the equipartition with
the emitted radiation field, we have $E_{\rm mag} \sim 25 r^{D-1}E_{\rm th}$ and the active regions
are  indeed magnetically dominated. In fact, the magnetic field intensity can be estimated as
\begin{equation}
B\simeq 1.6 \times 10^4 \tau_{\rm T}^{1/2} r^{(D-1.5)} M_7^{-1/2}
 \left(\frac{b}{30}\right)^{1/2}\; {\rm G}
.
\nonumber
\end{equation} 
Furthermore, we have $v_{\rm A}/c \simeq 0.03 r^{D-1} b^{1/2}$. This translates into
$v_{\rm A}/c \simeq 0.1 r^{2(D-1)/3} (\ln{R_{\rm m}})^{1/3}$, in the case of Petschek-type reconnection (see section
\ref{structure}). It is evident that 
the slow Sweet-Parker reconnection model is not compatible with the above scenario, because it would lead to
 too large a magnetic field accumulated in the corona. Our adopted value of the parameter $b$, is instead 
more appropriate for a relatively fast reconnection, as in the Petschek-type model.
In principle the value of such a parameter could be inferred from temporal studies, if we independently knew 
 the physical characteristic
size of an emitting region, as its value affects primarily the timescale of the 
individual flares.  

To conclude, as already discussed in section \ref{model}, we would like to remark on the fact that 
 synchrotron radiation is hardly important as source
of seed photons for Comptonization in AGN, but for the hardest, compact short lived flares. However, direct
synchrotron light at frequencies near the self-absorption cut-off 
could be observed below the peak of the disc quasi-thermal hump 
in sources where the direct emission from the accretion disc is
strongly suppressed (i.e. for low accretion rates, high fraction of power dissipated in the corona
and/or high relativistic bulk motion of the corona itself).

\subsection{Iron line variability}
The X-ray spectra of accreting black holes, and those of Seyfert 1s
in particular, often show
a clear sign of reflected radiation in the form of a prominent fluorescent iron
K$\alpha$ line. This spectral component is sensitive to the geometry
of the disc--corona system and can be then used to place more firm constraints on it.
Often such lines are Doppler and gravitationally broadened,
suggesting that the cold material within which
they are produced is moving at relativistic speeds in the deep potential well of the central
black hole (Fabian et al. 2000).
In the standard picture in which the reflection features are produced by the accretion disc
illuminated by the hard X-rays coming from  the active regions, the strong observed
continuum variability should drive the line variability with a short time-lag.

This simple picture, though, is challenged by the observations, which seem to
indicate that, although the iron line does indeed vary with the varying flux,
its flux is not correlated with the changes of the continuum intensity or of its slope (Vaughan
\& Edelson 2001).

In the framework of the thundercloud model, such complicated behaviour can be explained
by the fact that the observed line at any time is the super-position of different features
produced by active regions illuminating the disc with different spectral shape and intensity. The
ionization properties of the surface of the accretion disc, which in turn determine the line
characteristics, are strongly dependent on the illuminating continuum properties
\cite{nkk00,dn01,brf01,bal01}. Therefore the observed iron line is expected to be complex, 
and variable on a wide range
of timescales in a complicated way. 

Furthermore, as the coronal optical depth becomes larger
the reflection features are strongly suppressed by Compton scattering in the corona itself
(Matt et al. 1997; Petrucci et al. 2001b). This effect becomes more important the larger is the active region.
Analogously to what discussed in Malzac et al. (2001), we define a geometrical parameter that
determines the fraction of reflected luminosity that reenters an active region, 
$\mu_s=h/\sqrt{4r^2+h^2}$. Therefore, 
we can quantify the effect of an obscuring thundercloud 
by estimating the iron K$\alpha$ line luminosity produced by  any given 
illuminating spectrum of an active region in the following way:
\begin{equation}
\label{iron_line}
L_{K\alpha}=(L_{7-20}/2) a [\mu_s + (1-\mu_s)(1-P_{\rm sc}(\tau_{\rm T}))],
\end{equation}
where $L_{7-20}/2$ is the luminosity above the iron K$\alpha$ edge emitted towards
the disc and $P_{\rm sc}$ is the scattering probability in an active region (see Appendix A). 
A fraction $\mu_s$ of the reflected X-rays reaches the observed directly,
while a fraction $(1-\mu_s)P_{\rm sc}$ suffers
further Compton scattering in the corona and is therefore scattered out of the line, reducing
its absolute flux. We note here that the above expression may be in fact underestimating the amount of
cloud obscuration because the optical depth $\tau_{\rm T}$ 
that enters in the determination of $P_{\rm sc}$ 
is the one calculated from the center of an active region
 and is smaller than that seen by the photons
coming from the underlying disc.

\begin{figure}
%\vbox to110mm{\vfil
\psfig{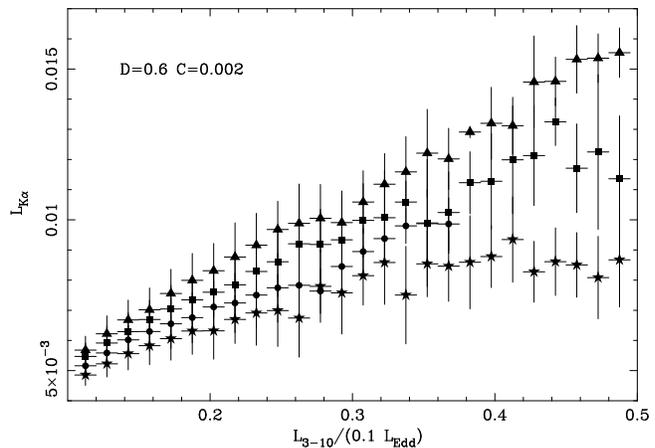}
\caption{The estimated iron line luminosity, as calculated in Eq. (\ref{iron_line}) as a function of the 
observed flux in the 3-10 keV band (in units of $0.1 L_{\rm Edd}$) 
for the simulation with $D=0.6$ and $C=0.002$.  
The points for all the time intervals have been grouped in luminosity
bins and the average line flux has been calculated. The error bars correspond to the weighted statistical 
uncertainty. 
Triangles corresponds to $\tau_{\rm T}=0.5$, squares to  $\tau_{\rm T}=1$,
circles to  $\tau_{\rm T}=1.5$ and stars to $\tau_{\rm T}=2$.} 
%\vfil}
\label{iron}
\end{figure}

In Figure \ref{iron} we show the estimated total Fe K$\alpha$ line luminosity 
produced by all the simultaneous active regions for a simulation with $C=0.001$ and $D=1$, 
plotted as a function of the total observed
3-10 keV luminosity. Clearly, in the case of high coronal optical 
depth ($\tau_{\rm T}\ga 1.5$), although still mildly correlated with the observed continuum at low
 luminosity (small flares and small timescales), 
the line variability is strongly suppressed at high luminosity (larger flares and longer timescales), 
to the extent that for a variation of a factor of three
in the 3-10 keV flux, the line flux may increases only by about $50\%$.

In general, even for less dramatic cases, we expect the line equivalent width $W_{K\alpha}$ 
to decrease with the increasing flux.
For example, in the simulation presented in Fig.~\ref{iron}, for $\tau=1.5$, we have
$W_{K\alpha} \sim 300$eV for $L_{3-10}\simeq 0.1$ and $W_{K\alpha} \sim 200$eV for $L_{3-10}\simeq 0.3$.
Such an high value for the unattenuated iron line equivalent width (compared to the expected value of 
$\sim 100$ eV; George \& Fabian 1991)
can be consistent with a
static corona model if the iron is overabundant with respect to the solar value, 
as indeed has been shown to be required to explain the reflection features of MCG--6-30-15 \cite{lee99}.

A full simulation of the temporal behaviour of the iron line, which is clearly beyond the scope
of this work, will involve a full treatment of the ionization state of the accretion disc atmosphere
  below different active regions (with different illuminating spectra) {\it and} of the obscuring effects
of the corona itself. Similarly, the effect of Comptonization in the corona on the shape and
intensity of the reflection hump may reveal themselves in time-resolved spectral analysis of the hard X-ray
emission, though this may prove a difficult task for the existing instruments capabilities \cite{pet01b}.

\section{Conclusion}
\label{conclusion}
We have presented a model for spectral and temporal variability of the X-ray emission from
accretion disc coronae around black holes.
We have assumed thermal Comptonization as the main radiation mechanism to produce the observed 
X-ray continuum. The cooling radiation comes either from the intrinsic dissipation of gravitational 
energy in the underlying accretion disc, or from the reprocessed radiation in the cold disc or from
synchrotron radiation produced locally in the magnetically dominated corona. 
For the case of a typical AGN of $10^7 M_{\odot}$, accreting at ten percent of the Eddington ratio and
for which most of the power is dissipated in the corona ($f_{\rm H}\ga 0.7$), we have shown that the 
reprocessed radiation is usually the main source of soft photons but from the case of more compact and hot
active regions, where the contribution from synchrotron radiation is 
important. Nonetheless, the magnetic field in the corona acts as the main energy reservoir
and dominates the local energy budget.

The basics geometric properties of the coronal
flow assumed in our model and, we believe, required
to explain the observed spectral and temporal variability of AGN are the
following:
\begin{itemize}
\item{The corona must not be uniform, but structured and heated intermittently
(Merloni \& Fabian 2001);}
\item{The fundamental heating event, a flare likely caused by magnetic reconnection,
must be compact, with typical size comparable to the accretion
disc thickness;}
\item{The height of the reconnection site must be at least an order of magnitude larger than
its size;}
\item{The spatial and temporal distribution of the flares are not random, but proceed in
correlated trains of events in an avalanche fashion;}
\item{The size of the avalanches determines the size of the active regions and their luminosity, and are
distributed as a power-law.}
\end{itemize}

We have simulated X-ray lightcurves of AGN and studied the correlation between the photon 
index and the X-ray luminosity. By comparing our simulations with  observation of the 
best studied Seyfert 1 galaxy MGC--6-30-15 we conclude that the model is able to reproduce
the observed $\Gamma$-luminosity correlation and to put some constraints on the model parameter.
In particular, to explain the trend observed in the 1997 {\it RXTE} observations analysed by Vaughan \& Edelson (2001)
we need a fairly high coronal optical depth ($\tau_{\rm T}\ga 1.5$), a small {\it average} 
covering fraction ($\sim 10^{-3}$) and
a scaling index for the active region luminosity $D \sim 1$. Furthermore our model 
naturally explains the linear correlation between luminosity and RMS variability 
(at different epochs in the same source), recently discovered by Uttley \& McHardy (2001) in a number
of accretion-powered compact objects.

Our model does not include other important geometrical or dynamical parameters that may vary on timescales
much longer than that over which the corona is heated (which is close to the dynamical one) 
and may indeed be relevant to determine the average long-term spectral
properties of these sources. In fact, our model does not reproduce the observed correlation
between the spectral index and the reflection features \cite{zls99,lz01}, and no `static' corona 
does, as demonstrated by Malzac et al. (2001). 
If the corona instead is in stable relativistic bulk
motion (i.e. is the base of a jet/outflow), and the outflow velocity changes with time, or from source to source,
then  we expect the average spectral index to correlate with the amount of reflection
\cite{bel99a,mal00}, as observed. Then our model would be applicable as it is only to the sources
with the softest spectra (and indeed the spectral parameters of MCG--6-30-15 correspond
to negligible bulk coronal velocity in the model of Beloborodov 1999a).

On the other hand, such a correlation has not yet been proved to hold
on the short timescales we focused our attention on. The short-time variability 
should bear information on some fundamental properties of any accreting system in which
an optically thick and geometrically thin disc coexist with a hot comptonizing medium
heated by dissipation of a structured magnetic field. 
A dynamic corona, where the reconnection sites 
moves with relativistic velocities and generate an outflow whose global properties 
evolve on  timescales much longer than the dynamical one, is a straightforward generalization of
the thundercloud model we have presented here, and should be consider in order to better
constrain the geometry and dynamics of the inner accretion flow around black holes.

Thus, detailed analysis of time-averaged spectra will help determining the value of the parameter that
is ultimately responsible for the long-term variability 
and therefore for the $\Gamma$-R correlation (such as the bulk outflow velocity, for example), while 
spectral and temporal variability on short timescales will ultimately probe the physics of coronal heating
and magnetic reconnection, providing a test for the model we have illustrated in this work. 

\section*{Acknowledgments}
We thank the referee, Dr. Andrzej Zdziarski, for the useful comments and suggestions. 
We thank Simon Vaughan for the MGC--6-30-15 data and for fruitful discussion, together with
David Shih, Kazushi Iwasawa and David Ballantyne.  
This work was done in the research network
``Accretion onto black holes, compact stars and protostars"
funded by the European Commission under contract number
ERBFMRX-CT98-0195'. AM and ACF thank the PPARC and the Royal Society
for support, respectively.

\appendix
\section{The X-ray spectrum}

We calculate the self-absorbed synchrotron luminosity
using the expression given in Wardzi\'nski \& Zdziarski (2000).
The intrinsic disc luminosity is simply $L_{\rm int}=\dot m (1-f_{\rm H}) L_{\rm Edd}$, 
where $\dot m$ is the accretion rate in units of the Eddington one, $ f_{\rm H}$ is
the fraction of the accretion power dissipated in the corona
(Svensson \& Zdziarski 1994) and $L_{\rm Edd} = 1.3 \times 10^{38} M/M_{\odot}$
is the Eddington luminosity.
Finally, the reprocessed luminosity, $L_{\rm
rep}(r)$, is calculated from Eq. (\ref{lrep}). For the two corresponding
thermal soft photon fluxes from the disc $F_{\rm bb,i}$ and $F_{\rm bb,r}$, we
assume black-body spectra, with temperatures $\Theta_{\rm i}$ and
$\Theta_{\rm r}$, respectively. They are given by:
\begin{eqnarray}
\Theta_{\rm i}&=&\left(\frac{k}{m_{\rm e}c^2}\right)
\left(\frac{L_{\rm int}}
{\sigma_{\rm B} 4 \pi R_{\rm disc}^2}\right)^{1/4} \\
\Theta_{\rm r}&=&\left(\frac{k}{m_{\rm e}c^2}\right)
\left(\frac{L_{\rm rep}(r)} {\sigma_{\rm B} 4 \pi R_{\rm
hot}^2}\right)^{1/4}.
\end{eqnarray}

\begin{figure}
%\vbox to140mm{\vfil
\psfig{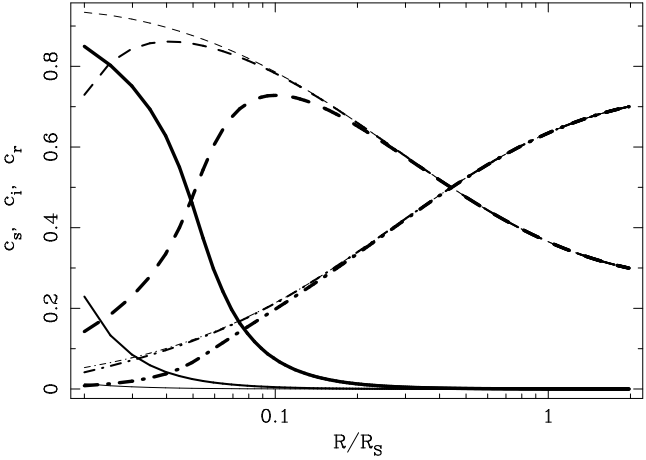}
\caption{The relative contribution to the cooling luminosity from synchrotron
radiation ($c_{\rm s}$, solid lines), intrinsic dissipation in the 
accretion disc ($c_{\rm i}$, dashed lines)
and from reprocessed radiation ($c_{\rm r}$, dot-dashed lines), as functions of an active region size $r$ for
$D=1$ and a covering fraction of $10^{-3}$.
The energy densities are calculated for a fixed height above the disc $H=1R_{\rm S}$ in the case of
 an AGN with $M_{\rm BH}=10^7 M_{\odot}$ accreting with $\dot m=0.1$, $f_{\rm H}=0.7$ and
 $\epsilon_{\rm M}=0.1$. The thickness of the lines indicates the optical depth: in order of 
decreasing thickness we have $\tau_{\rm T}=0.5,1,2$.
 Clearly synchrotron radiation is important only for the more compact sources with the
lowest optical depth, which are hotter and produce harder spectra, as shown also in
 Figure 1. Reprocessed radiation energy density, instead, 
always dominates as a source of soft photons for Comptonization for large enough active regions. Such 
behaviour is common for all values of the parameters; as a rule, the reprocessed radiation is more important
for smaller covering fraction and higher values of the fraction of power dissipated in the corona $f_{\rm H}$.} 
%\vfil} 
\label{fig_coeff}
\end{figure}

Comparing the energy densities of the three photon fields inside
each active region, we determine the coefficients $c_{\rm s}$,
$c_{\rm i}$ and $c_{\rm r}$, with $c_{\rm s}+c_{\rm
i}+c_{\rm r}=1$, that fix the relative contribution of synchrotron, intrinsic and
reprocessed radiation 
soft photons sources, respectively. This is illustrated in Fig. \ref{fig_coeff} for our fiducial
model parameters 
and $\epsilon_{\rm M}=0.1$. It is evident that for $r \ga 0.1 h$ the contribution of 
synchrotron radiation to the cooling luminosity is always negligible, 
in particular for the high optical depth (lower temperature) cases.

The luminosity due to
the scattered photons can be approximated as a sum of a cut-off
power-law and Wien component (Wardzi\'nski \& Zdziarski 2000),
\begin{equation}
L_{\rm C}=N_{\rm P}\left(\frac{x}{\Theta}\right)^{-\alpha} \exp^{-x/\Theta}+
N_{\rm W}\left(\frac{x}{\Theta}\right)^{3} \exp^{-x/\Theta},
\end{equation}
where $x=h\nu/m_{\rm e}c^2$ is a dimensionless photon energy, $N_{\rm P}$ and
$N_{\rm W}$ are the normalizations of the power-law and Wien tail, respectively,
and are related by the ratio (Zdziarski 1985)
 \begin{equation}
\frac{N_{\rm W}}{N_{\rm P}}=\frac{\Gamma_{\rm E}(\alpha)}{\Gamma_{\rm E}(2\alpha+3)}P_{\rm sc},
\end{equation}
where $\Gamma_{\rm E}(x)$ is Euler's Gamma function.

The scattering probability averaged over the source volume, $P_{\rm sc}$,
depends on the coronal optical depth only, and determine the emerging spectral index:
\begin{equation}
\alpha=-\frac{\ln P_{\rm sc}}{\ln A}.
\end{equation}
In spherical geometry (Osterbrock 1974),
\begin{equation}
P_{\rm sc}=1-\frac{3}{8\tau_{\rm T}^3}\left[2\tau_{\rm T}^2-1+\exp^{-2\tau_{\rm T}}
(2\tau_{\rm T}+1)\right].
\end{equation}

Finally, we have for the power-law normalization
\begin{eqnarray}
\lefteqn{ N_{\rm P} =  4 \pi r^2 \phi \left[
   c_{\rm i} F_{\rm bb,i}(\Theta_{\rm i})
\left(\frac{2.78 \Theta_{\rm i}}{\Theta}\right)^{\alpha}+\right. } \nonumber\\
\lefteqn{\left. \qquad c_{\rm r} F_{\rm bb,r}(\Theta_{\rm r}) \left(\frac{2.78
\Theta_{\rm r}}{\Theta}\right)^{\alpha} + c_{\rm s} F_{\rm s}
\left(\frac{x_{\rm t}}{\Theta}\right)^{\alpha} \right] }
\end{eqnarray}
where the synchrotron flux $F_{\rm s}=2 \pi m_{\rm e}c^3 \Theta
x_{\rm t}^2/ \lambda_{\rm c}^3$ and $x_{\rm t}$ is the turnover
energy (in units of $m_{\rm e} c^2$) of the self-absorbed
synchrotron radiation, that has been calculated as in Di Matteo
Celotti \& Fabian (1997). The factor  $\phi \approx
(1+4\Theta^2)/(1+40 \Theta^2)$ approximates the relative
normalization of the power-law extrapolation and the soft photon
input peak flux, and is derived form a phenomenological analysis
(Zdziarski 1986).

The coronal temperature, and consequently the slope of the
comptonized continuum emerging from each active region, are
calculated  self-consistently solving for $L_{\rm C}=L(r)=c_1
\dot m f_{\rm H} L_{\rm Edd} r^D$.

\bsp

\label{lastpage}

\end{document}